\documentclass{webofc}
\usepackage[varg]{txfonts}   
\usepackage{hyperref}
\usepackage{textcomp}
\usepackage{gensymb}
\usepackage{amssymb, amsmath}

\begin{document}

\title{Dust Emission in Galaxies at Millimeter Wavelengths}
\subtitle{Cooling of star forming regions in NGC6946}

\author{\firstname{G.}~\lastname{Ejlali}\inst{\ref{IPM}}\fnsep\thanks{\email{gejlali@ipm.ir}}
\and  \firstname{R.}~\lastname{Adam} \inst{\ref{LLR}}
\and  \firstname{P.}~\lastname{Ade} \inst{\ref{Cardiff}}
\and  \firstname{H.}~\lastname{Ajeddig} \inst{\ref{CEA}}
\and  \firstname{P.}~\lastname{Andr\'e} \inst{\ref{CEA}}
\and  \firstname{E.}~\lastname{Artis} \inst{\ref{LPSC}}
\and  \firstname{H.}~\lastname{Aussel} \inst{\ref{CEA}}
\and  \firstname{A.}~\lastname{Beelen} \inst{\ref{IAS}}
\and  \firstname{A.}~\lastname{Beno\^it} \inst{\ref{Neel}}
\and  \firstname{S.}~\lastname{Berta} \inst{\ref{IRAMF}}
\and  \firstname{L.}~\lastname{Bing} \inst{\ref{LAM}}
\and  \firstname{O.}~\lastname{Bourrion} \inst{\ref{LPSC}}
\and  \firstname{M.}~\lastname{Calvo} \inst{\ref{Neel}}
\and  \firstname{A.}~\lastname{Catalano} \inst{\ref{LPSC}}
\and  \firstname{I.}~\lastname{de Looze} \inst{\ref{Belguim},\ref{London}}
\and  \firstname{M.}~\lastname{De~Petris} \inst{\ref{Roma}}
\and  \firstname{F.-X.}~\lastname{D\'esert} \inst{\ref{IPAG}}
\and  \firstname{S.}~\lastname{Doyle} \inst{\ref{Cardiff}}
\and  \firstname{E.~F.~C.}~\lastname{Driessen} \inst{\ref{IRAMF}}
\and  \firstname{M.}~\lastname{Galametz} \inst{\ref{CEA}}
\and  \firstname{F.}~\lastname{Galliano} \inst{\ref{CEA}}
\and  \firstname{A.}~\lastname{Gomez} \inst{\ref{CAB}}
\and  \firstname{J.}~\lastname{Goupy} \inst{\ref{Neel}}
\and  \firstname{A. P.}~\lastname{Jones} \inst{\ref{IAS}}
\and  \firstname{A.}~\lastname{Hughes} \inst{\ref{Toulouse}}
\and  \firstname{S.}~\lastname{Katsioli} \inst{\ref{Athens},\ref{Athens2}}
\and  \firstname{F.}~\lastname{K\'eruzor\'e} \inst{\ref{LPSC}}
\and  \firstname{C.}~\lastname{Kramer} \inst{\ref{IRAME}}
\and  \firstname{B.}~\lastname{Ladjelate} \inst{\ref{IRAME}}
\and  \firstname{G.}~\lastname{Lagache} \inst{\ref{LAM}}
\and  \firstname{S.}~\lastname{Leclercq} \inst{\ref{IRAMF}}
\and  \firstname{J.-F.}~\lastname{Lestrade} \inst{\ref{LERMA}}
\and  \firstname{J.-F.}~\lastname{Mac\'ias-P\'erez} \inst{\ref{LPSC}}
\and  \firstname{S. C.}~\lastname{Madden} \inst{\ref{CEA}}
\and  \firstname{A.}~\lastname{Maury} \inst{\ref{CEA}}
\and  \firstname{P.}~\lastname{Mauskopf} \inst{\ref{Cardiff},\ref{Arizona}}
\and  \firstname{F.}~\lastname{Mayet} \inst{\ref{LPSC}}
\and  \firstname{A.}~\lastname{Monfardini} \inst{\ref{Neel}}
\and  \firstname{M.}~\lastname{Mu\~noz-Echeverr\'ia} \inst{\ref{LPSC}}
\and  \firstname{A.}~\lastname{Nersesian} \inst{\ref{Athens},\ref{Belguim}}
\and  \firstname{L.}~\lastname{Perotto} \inst{\ref{LPSC}}
\and  \firstname{G.}~\lastname{Pisano} \inst{\ref{Cardiff}}
\and  \firstname{N.}~\lastname{Ponthieu} \inst{\ref{IPAG}}
\and  \firstname{V.}~\lastname{Rev\'eret} \inst{\ref{CEA}}
\and  \firstname{A.~J.}~\lastname{Rigby} \inst{\ref{Cardiff}}
\and  \firstname{A.}~\lastname{Ritacco} \inst{\ref{IAS}, \ref{ENS}}
\and  \firstname{C.}~\lastname{Romero} \inst{\ref{Pennsylvanie}}
\and  \firstname{H.}~\lastname{Roussel} \inst{\ref{IAP}}
\and  \firstname{F.}~\lastname{Ruppin} \inst{\ref{MIT}}
\and  \firstname{K.}~\lastname{Schuster} \inst{\ref{IRAMF}}
\and  \firstname{S.}~\lastname{Shu} \inst{\ref{Caltech}}
\and  \firstname{A.}~\lastname{Sievers} \inst{\ref{IRAME}}
\and  \firstname{M. W. S. L.}~\lastname{Smith}\inst{\ref{Cardiff}}
\and  \firstname{F. S.}~\lastname{Tabatabaei}\inst{\ref{IPM},\ref{Canarias}}
\and  \firstname{C.}~\lastname{Tucker} \inst{\ref{Cardiff}}
\and  \firstname{E. M.}~\lastname{Xilouris}\inst{\ref{Athens}}
\and  \firstname{R.}~\lastname{Zylka} \inst{\ref{IRAMF}}
}
  
\institute{
  	 School of Astronomy, Institute for Research in Fundamental Sciences (IPM), Larak Garden, 19395-5531 Tehran, Iran 
  	\label{IPM}
  	\and
	LLR (Laboratoire Leprince-Ringuet), CNRS, École Polytechnique, Institut Polytechnique de Paris, Palaiseau, France
	\label{LLR}
	\and
	School of Physics and Astronomy, Cardiff University, Queen’s Buildings, The Parade, Cardiff, CF24 3AA, UK 
	\label{Cardiff}
	\and
	AIM, CEA, CNRS, Universit\'e Paris-Saclay, Universit\'e Paris Diderot, Sorbonne Paris Cit\'e, 91191 Gif-sur-Yvette, France
	\label{CEA}
	\and
	Univ. Grenoble Alpes, CNRS, Grenoble INP, LPSC-IN2P3, 38000 Grenoble, France
	\label{LPSC}
	\and
	Institut d'Astrophysique Spatiale (IAS), CNRS, Universit\'e Paris Sud, Orsay, France
	\label{IAS}
	\and
	Institut N\'eel, CNRS, Universit\'e Grenoble Alpes, France
	\label{Neel}
	\and
	Institut de RadioAstronomie Millim\'etrique (IRAM), Grenoble, France
	\label{IRAMF}
	\and
	Aix Marseille Univ, CNRS, CNES, LAM, Marseille, France
	\label{LAM}
	\and 
	Dipartimento di Fisica, Sapienza Universit\`a di Roma, Piazzale Aldo Moro 5, I-00185 Roma, Italy
	\label{Roma}
	\and
	Univ. Grenoble Alpes, CNRS, IPAG, 38000 Grenoble, France 
	\label{IPAG}
	\and
	Centro de Astrobiolog\'ia (CSIC-INTA), Torrej\'on de Ardoz, 28850 Madrid, Spain
	\label{CAB}
	\and
	Universit\'e de Toulouse, UPS-OMP,F-31028 Toulouse, France; CNRS, IRAP, Av. du Colonel Roche BP 44346, F-31028 Toulouse cedex 4, France
	\label{Toulouse}
	\and  
	Instituto de Radioastronom\'ia Milim\'etrica (IRAM), Granada, Spain
	\label{IRAME}
	\and
	National Observatory of Athens, Institute for Astronomy, Astrophysics, Space Applications and Remote Sensing, Ioannou Metaxa and Vasileos Pavlou GR-15236, Athens, Greece
	\label{Athens}
	\and
	Department of Astrophysics, Astronomy \& Mechanics, Faculty of Physics,
	University of Athens, Panepistimiopolis, GR-15784 Zografos, Athens, Greece
	\label{Athens2}
	\and
	Sterrenkundig Observatorium Universiteit Gent, Krijgslaan 281 S9, B-9000 Gent, Belgium
	\label{Belguim}
	\and
	Department of Physics and Astronomy, University College London, Gower Street, London, WC1E 6BT, UK
	\label{London}
	\and 
	LERMA, Observatoire de Paris, PSL Research University, CNRS, Sorbonne Universit\'e, UPMC, 75014 Paris, France  
	\label{LERMA}
	\and
	School of Earth and Space Exploration and Department of Physics, Arizona State University, Tempe, AZ 85287, USA
	\label{Arizona}
	\and 
	Laboratoire de Physique de l’\'Ecole Normale Sup\'erieure, ENS, PSL Research University, CNRS, Sorbonne Universit\'e, Universit\'e de Paris, 75005 Paris, France 
	\label{ENS}
	\and
	Department of Physics and Astronomy, University of Pennsylvania, 209 South 33rd Street, Philadelphia, PA, 19104, USA
	\label{Pennsylvanie}
	\and 
	Institut d'Astrophysique de Paris, Sorbonne Université, CNRS (UMR7095), 75014 Paris, France
	\label{IAP}
	\and 
	Instituto de Astrofísica de Canarias, Vía L’actea S/N, 38205 La Laguna, Spain
	\label{Canarias}
	\and
	Kavli Institute for Astrophysics and Space Research, Massachusetts Institute of Technology, Cambridge, MA 02139, USA
	\label{MIT}
	\and
	Caltech, Pasadena, CA 91125, USA
	\label{Caltech}
}  
  
\abstract{Interstellar dust plays an important role in the formation of molecular gas and the heating and cooling of the interstellar medium. The spatial distribution of the mm-wavelength dust emission from galaxies is largely unexplored. The NIKA2 Guaranteed Time Project IMEGIN (Interpreting the Millimeter Emission of Galaxies with IRAM and NIKA2) has recently mapped the mm emission in the grand design spiral galaxy NGC6946. By subtracting the contributions from the free-free, synchrotron, and CO line emission, we map the distribution of the pure dust emission at $1.15mm$ and $2mm$. Separating the arm/interarm regions, we find a dominant $2mm$ emission from interarms indicating the significant role of the general interstellar radiation field in heating the cold dust. Finally, we present maps of the dust mass, temperature, and emissivity index using the Bayesian MCMC modeling of the spectral energy distribution in NGC6946.}

\maketitle

\section{Introduction}
\label{intro}
The IMEGIN project (PI: Suzanne Madden) is a NIKA2 guaranteed time large project (200 hours) on the 30m IRAM telescope, studying the interstellar medium (ISM) of 22 nearby galaxies. The NIKA2 camera, with a $6.5'$ field of view and resolution of $11.1''$ at $1.15mm$ and $17.6''$ at $2mm$~(\cite{NIKA2-performance}) brings us well-resolved, ($\sim kpc$ scale) studies to map full galaxies at these wavelengths for the first time. Here we present the results of a pilot study of the nearby ($D = 6.8Mpc$) grand-design spiral NGC6946, observed almost face-on ($i = 38^\circ$), with a wealth of complementary millimeter, radio, atomic observations as well as molecular lines.\par

\section{Observations}
\label{obs}
\begin{figure*}
	\centering
	\includegraphics[width=0.37\textwidth]{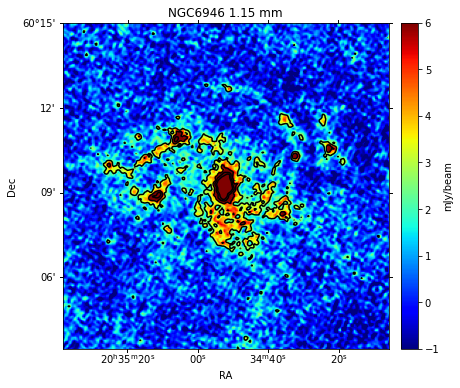}
	\includegraphics[width=0.37\textwidth]{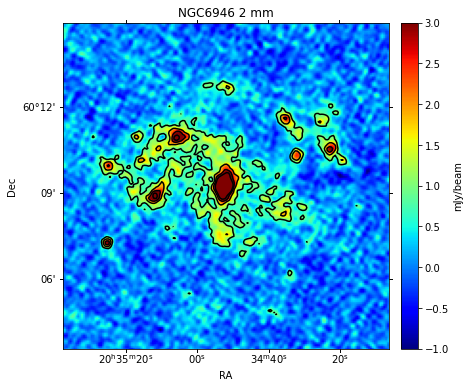}
	\caption{Observed maps of NGC6946 at $1.15 mm$  (\textit{left}) and $2 mm$ (\textit{right}) at angular resolution of $12''$ and $18''$ respectively.}
	\label{observedmaps} 
\end{figure*}

NGC6946 was observed in 2020 and 2021 for a total of 23 hours, reaching rms values of of $0.8 mJy/beam$ at $1.15 mm$ ($3''$ pixel) and $0.27  mJy/beam$ at $2 mm$ ($4''$ pixel), with $5\%$ calibration error~\cite{NIKA2-performance}. The scans were reduced and maps reconstructed using PIIC/GILDAS software\footnote{\url{http://www.iram.es/IRAMES/mainWiki/PIIC}} (Figure~\ref{observedmaps}). 

\section{Emission at millimeter wavelengths}
\label{emission}
Continuum emission at mm wavelengths consists of three main components, 1) thermal emission from dust, 2) thermal free-free emission from ionized gas, and 3) nonthermal synchrotron emission from cosmic rays propagating through the magnetic field~\cite{Tabatabaei2013b}. Apart from these, continuum emission might be affected by line emission of molecular gas, such as CO. \par

We use the thermal and nonthermal radio maps of NGC6946 that have been constructed from the H$\alpha$ recombination line observations~\cite{Tabatabaei2013a}. For pixels with flux values more than $3\sigma$ rms, we find that $5.5\%$ and $28.8\%$ of integrated flux at $1.15 mm$ and $2 mm$ is due to the radio continuum components. \par

Next, we take into account the contamination by molecular lines, as the NIKA2 $1.15mm$ band includes emission from the CO(2-1) line. We use the CO(2-1) map observed at IRAM as a part of HERACLES project~\cite{Leroy2009}. We subtract the CO(2-1) emission from the observed emission at $1.15mm$ following formulas explained in~\cite{Drabek2012}, which uses the transmission function of NIKA2~\cite{NIKA2-performance} to convert spectral intensity to pseudo-continuum units. The CO(2-1) emission accounts for $14.1\%$ of the integrated flux of NIKA2 $1.15 mm$ emission in NGC6946. \par

The maps of pure dust emission are shown in Figure 2. $80.4\%$ and $63.6\%$ of the integrated flux observed over NGC6946 in the NIKA2 $1.15mm$ and $2mm$ bands, respectively, corresponds to pure dust emission (for pixels with larger than $3\sigma$ rms observed flux).

\begin{figure*}
	\centering
	\includegraphics[width=0.37\textwidth]{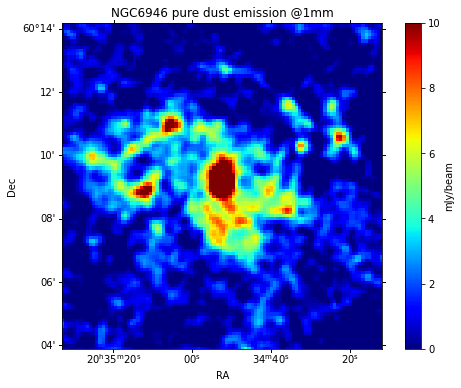}
	\includegraphics[width=0.37\textwidth]{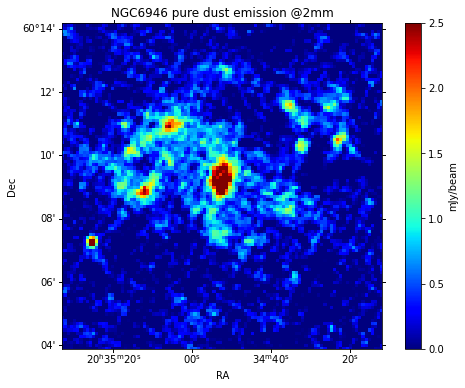}
	\caption{Pure dust emission at $1mm$ (\textit{left}) and $2mm$ (\textit{right}) at angular resolution of $18''$.}
	\label{puredust}
\end{figure*}

\section{Regional Analysis}
\label{regional}
\begin{figure*}
	\centering
	\includegraphics[width=0.20\textwidth]{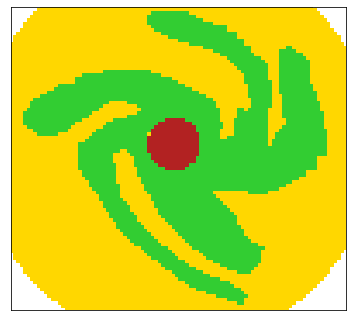}
	\includegraphics[width=0.70\textwidth]{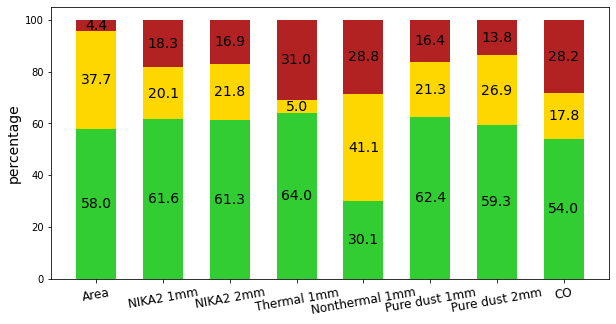}
	\caption{\textit{Left}: The mask we used to distinguish between central (red), arm (green) and interarm (yellow) regions in NGC6946. \textit{Right}: Percentage of contribution of each region to the indicated emissions.}
	\label{barchart}
\end{figure*}

To study the spatial properties of the ISM and dust in NGC6946, we investigate emission in the central, arm and interarm regions separately, making use of a modified version of the mask introduced in~\cite{Bigiel2020}. Our modification includes adding a small part of the western arm and a large portion of the interarm region, which is missed due to the limited field of view of SOFIA in \cite{Bigiel2020}. The relative contribution of  the different types of emission from each of these three regions in NGC6946 is shown in Figure~\ref{barchart}.

It can be seen in Figure~\ref{barchart} that the arm-to-interarm ratio is larger for thermal (free-free) radio emission than the nonthermal (synchrotron) emission. The synchrotron emission is caused by cosmic ray electrons that can propagate large distances in the interarm regions and into the diffuse disc. In addition, a slightly larger contribution of dust emitting at $1.15mm$ compared to $2mm$ in the arm regions is noticeable. The $2mm$ emission is a better tracer of cold dust, compared to the $1.15mm$ emission which can trace warmer dust. Main heating sources of warm dust are the star forming regions, which follow the spiral structure of the galaxy in the arms. On the other hand, colder dust can be heated by the interstellar radiation field (ISRF), which may be the main heating source in the diffuse disc and interarm region. These speculations are investigated further by modeling the dust properties spatially in the galaxy.\par

\section{Spectral Energy Distribution (SED) modeling}
\label{SED}
To study the physical properties of dust and their spatial variations across NGC6946, we model the SED of this galaxy from centimeter radio to FIR wavelengths. We use the far infrared \textit{Herschel} data and the VLA and Effelsberg data to fully constrain the SED (Table~\ref{alldata}). All of the maps were first brought to the same resolution of $18''$ using a Gaussian PSF and grid size of $6''$. We also subtracted the CO(2-1) emission from NIKA2 $1.15mm$ map before including in the SED modeling.\par
To model the radio continuum SED, we use a power law model (Equation~\ref{rceq}), which depends on three parameters, namely $A_{1}$ indicating the thermal free-free fraction, $A_{2}$ indicating the nonthermal synchrotron fraction, and $\alpha$ which is the synchrotron spectral index \cite{Tabatabaei2017}. 
\begin{equation}
	\label{rceq}
	S_{\nu}^{\text{RC}}=A_{1}\nu^{-0.1}+A_{2}\nu^{-\alpha}
\end{equation}
To model the thermal emission from dust, we assume a modified black body (MBB) model which constraints dust mass $M$, dust temperature $T$ and dust emissivity index $\beta$ as
\begin{equation}
	S_{\nu}^{\text{dust}}=\kappa_{0}\left(\frac{\nu}{\nu_{0}}\right)^{\beta}\times \frac{M}{D^{2}}\times B_{\nu}(T),	
\end{equation} 
where $B_{\nu}(T)$ is the Planck function and we adopt $\kappa_{0}=0.04 m^2/kgr$~\cite{Tabatabaei2013b}. Addition of these two models, $S_{\nu}^{RC}+	S_{\nu}^{dust}$, with a set of six free parameters $\{A_{1}, A_{2},\alpha,T,M,\beta\}$ construct our model.\par

To fit this model to our data (Table~\ref{alldata}), we use a Bayesian Markov Chain Monte Carlo (MCMC) approach making use of the emcee Python package~\cite{emcee}. This method generated a model library that encompasses all different combinations of model parameters. The likelihood function of each model in the library is an indicator of how close that model is to the observed data. The Bayesian MCMC approach then derives the posterior distribution of each parameter using a specified number of walkers in the parameter space. The median of the posterior probability distribution function is then used as the reported result.\par

\begin{figure*}
	\centering
	\includegraphics[width=0.7\textwidth]{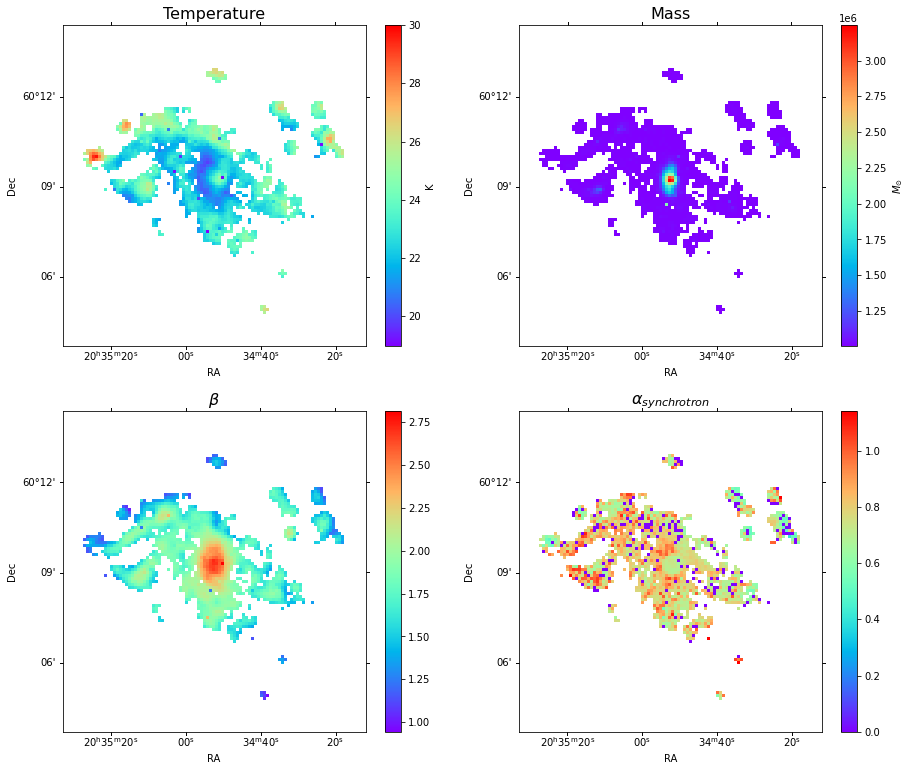}
	\caption{Maps of dust mass $M$, dust temperature $T$, dust emissivity index $\beta$ and synchrotron spectral index $\alpha$ as a result of SED modeling in each $6''$ pixel. The maps are at $18''$ resolution.}
	\label{SEDparams}
\end{figure*}

We fit our model to the observed flux densities of each pixel of the map and report the six free parameters for each pixel. The maps of the 4 main dust properties which we find through the SED modeling are demonstrated in Figure~\ref{SEDparams}.

\begin{table}
\centering
\caption{Summary of data used for resolved SED modeling.}
\label{alldata}
\small
\begin{tabular}{llll}
\hline
Wavelength & telescope & original resolution & reference \\\hline
$20cm$&VLA \& Effelsberg& 15"& \cite{Beck2007} \\
$6cm$&VLA \& Effelsberg& 15"& \cite{Beck2007}\\
$3.6cm$&VLA \& Effelsberg& 18"& \cite{Beck2007}\\
$1.15mm$& IRAM 30m NIKA2& 12"& This work \\
$2mm$& IRAM 30m NIKA2& 18"& This work \\
$70\mu m$& Herschel PACS& 6"& \cite{Kennicutt2011},\cite{Clark2018}\\
$100\mu m$& Herschel PACS& 8"& \cite{Kennicutt2011},\cite{Clark2018}\\
$160\mu m$& Herschel PACS& 12"& \cite{Kennicutt2011},\cite{Clark2018}\\
$250\mu m$& Herschel SPIRE& 18"& \cite{Kennicutt2011},\cite{Clark2018}\\ \hline
\end{tabular}
\end{table}

\section{Discussion}

We studied the correlation of dust emission with the CO(2-1) emission, as a tracer of the molecular gas. For this purpose, we only use pixels with flux values larger than  $3 \sigma$ RMS of both maps and separated by one resolution element. We find a strong correlation (Pearson coefficient = 0.978) between dust emission at $1.15 mm$ and CO(2-1) for pixels located in the central region of NGC6946. A similar trend has been reported in the central region of the Milky Way~\cite{Eden2020}. On the other hand, points belonging to arm and interarm region show a much weaker correlation (Pearson coefficient=0.600) between $1.15 mm$ emission of dust and molecular gas. This observation brings us to conclude that dust and molecular gas have a common heating source in the central region, whereas they are being heated by different heating sources in the disc. Both the diffuse interstellar radiation field and old stars play important roles in heating the dust in the disk, while they are not energetic enough to bring molecular gas to thermal equilibrium with dust. \par

We report an anti-correlation between $T$ and $\beta$ for galactocentric radii larger than $\sim1.5kpc$ in NGC6946. The fact that the anti-correlation breaks down for the inner $\sim1.5kpc$ brings us to conclude that this is not caused by degeneracies in our model and both $\beta$ and $T$ are well constrained.\par

\section*{Acknowledgements}
We would like to thank the IRAM staff for their support during the campaigns. The NIKA2 dilution cryostat has been designed and built at the Institut N\'eel. In particular, we acknowledge the crucial contribution of the Cryogenics Group, and in particular Gregory Garde, Henri Rodenas, Jean Paul Leggeri, Philippe Camus. This work has been partially funded by the Foundation Nanoscience Grenoble and the LabEx FOCUS ANR-11-LABX-0013. This work is supported by the French National Research Agency under the contracts "MKIDS", "NIKA" and ANR-15-CE31-0017 and in the framework of the "Investissements d’avenir” program (ANR-15-IDEX-02). This work has benefited from the support of the European Research Council Advanced Grant ORISTARS under the European Union's Seventh Framework Programme (Grant Agreement no. 291294).
This work was supported by the Programme National “Physique et Chimie du Milieu Interstellaire” (PCMI) of CNRS/INSU with INC/INP co-funded by CEA and CNES. F.R. acknowledges financial supports provided by NASA through SAO Award Number SV2-82023 issued by the Chandra X-Ray Observatory Center, which is operated by the Smithsonian Astrophysical Observatory for and on behalf of NASA under contract NAS8-03060.

\end{document}